\documentclass{PoS}
\usepackage{epsfig}
\usepackage{latexsym}
\usepackage{amsmath}
\usepackage{amsfonts}
\PoS{PoS(LAT2005)237}

\title{Wilson fermions quark bilinears to three loops}

\ShortTitle{Wilson fermions quark bilinears to three loops}

\author{\speaker{Francesco Di Renzo}\\
        University of Parma and I.N.F.N.\\
        E-mail: \email{direnzo@fis.unipr.it}}
        
\author{Andrea Mantovi\\
        University of Parma and I.N.F.N.\\
        E-mail: \email{mantovi@fis.unipr.it}}        
        
\author{Vincenzo Miccio\thanks{Current address: I.N.F.N. Milano.}\\
        University of Parma and I.N.F.N.\\
        E-mail: \email{vincenzo.miccio@mib.infn.it}}      

\author{Christian Torrero\\
        University of Parma and I.N.F.N.\\
        E-mail: \email{torrero@fis.unipr.it}}
        
\author{Luigi Scorzato\\
        Humboldt University, Berlin\\
        E-mail: \email{Luigi.Scorzato@physik.hu-berlin.de}}

\abstract{Quark currents renormalization constants can in principle be safely 
computed in lattice perturbation theory. In practice, traditional lattice perturbative 
computations are quite cumbersome, so that so far only the first loop results were available. 
By making use of Numerical Stochastic Perturbation Theory we reached three (and with less statistical precision 
even four) loops, both in quenched and in unquenched theory. Convergence properties of the series can be assessed 
and comparison with non perturbative results (where available) can be made: high loops computations of renormalization constants can be a valuable tool for lattice QCD.}

\FullConference{XXIIIrd International Symposium on Lattice Field Theory\\
                25-30 July 2005\\
                Trinity College, Dublin, Ireland}

\begin{document}

%
\section{Introduction}
Despite the fact that logarithmic divergent renormalization constants like those for quark bilinears could be in principle safely computed in Lattice Perturbation Theory (LPT), one usually tries to compute them non-perturbatively,  relying on some intermediate scheme. Popular choices are \emph{RI'-MOM} \cite{SPQR} and \emph{SF} \cite{Alpha}. Drawbacks of traditional perturbative computations are well known. LPT is hard and as a result computations of renormalization constants are usually at one loop, with second order computations just now making their entrance\footnote{See for example \cite{Quentin} at this conference.}. Due to the bad convergence properties of LPT, large use is made of so-called Boosted Perturbation Theory (BPT), often in the Tadpole-Improved (TI) variant. By making use of the Numerical Stochastic Perturbation Theory (NSPT) method\footnote{The method is reviewed in \cite{NSPTfull} and new technical details were also presented at this conference \cite{EnzoLAT05}.}, we are in a position to perform quite high orders computations: this means three (and even four) loops for quark bilinears. By making use of BPT, we are in a position to assess convergence properties and truncation errors of the series and to make contact with non perturbative results, where available\footnote{This is the case for the quantities at hand: see \cite{CecVittLAT05} at this conference.}. This is relevant, since both perturbative and non perturbative computations rely on the same assumptions, which strictly speaking are only proved in the perturbative framework. 

%
\section{Computational setup}

We performed our computations for Wilson gauge action and (unimproved, at the moment) Wilson fermion action ($r=1$). The quenched computations have been performed on lattice sizes of both $32^4$ and $16^4$. Unquenched computations are at the moment\footnote{$n_f=4$ computations are also on their way.} for $n_f = 2,3$ on $32^4$. The bigger lattice size fits well on our \emph{APEmille} crate, while standard PC's are enough for the smaller size. In the following we will focus on the $n_f=2$ results. Configurations have been stored in order to perform many other computations. 

We work in the \emph{RI'-MOM} scheme \cite{SPQR}. Quark bilinears operators are computed between (off shell) quark states of momentum $p$ and then amputated to get $\Gamma$-functions
\begin{equation}
	\langle p | \, \overline{\psi} \Gamma \psi \, | p \rangle \, = \, G(p) \;\;
\rightarrow \;\; \Gamma(p) \, = \, S^{-1}(p) G(p) S^{-1}(p).
\end{equation}
By making use of a convenient projector $P_O$ one then projects on the tree level structure
\begin{equation}
	O(p) = \mbox{Tr}\left(\hat{P_O} \Gamma(p)\right)
\end{equation}
to obtain the operators in terms of which the renormalization conditions are given
\begin{equation} \label{master}
	Z_O \, Z_q^{-1} \, O(p) \Big|_{p^2 = \mu^2} \, = \, 1.
\end{equation}
$Z_q$ is the quark field renormalization constants defined via
\begin{equation}
	Z_q \, = \, - i \frac{1}{12} \frac{\mbox{Tr}(\not p S^{-1})}{p^2}\Big|_{p^2 = \mu^2}.
\end{equation}
Renormalization conditions are given in the zero quark mass limit in order to get a mass independent scheme. We note with this respect that we are in a position to stay at zero mass by plugging in the Dirac operator the convenient (critical mass) counterterms. These are analytically known at one and two loops \cite{PanaPelo}, while third loop has been computed in \cite{NSPTfull}. Going to four loop, we obtained as a byproduct the corresponding new order for the critical mass\footnote{A quite surprising result is that resumming the critical mass at four loop in a convenient BPT scheme one obtains a result which is not so far from non perturbative determinations.}. 

The gauge is fixed to Landau, which determines the one loop quark field anomalous dimension to vanish. 

One should keep in mind that the \emph{RI'-MOM} scheme is defined in infinite volume, while we are of course forced to finite volume approximations: as we will see, care is to be taken of this point. 

In a traditional perturbative computation fixing the divergences (\emph{i.e.} the $\log$'s) is the \emph{easy} part of the job, while obtaining finite parts is usually much harder. In NSPT it is just the other way around. We actually take the anomalous dimensions $\gamma$'s for granted. In \emph{RI'-MOM} they are known to three loop order \cite{JG}: notice that is the reason why we stop at third order. Only for RG-invariant quantities (like the ratio $Z_p/Z_s$ on which we will focus in the following) we have almost no conceptually severe limitations. 

In the end, at a given loop $L$ this is what one expects for the coefficient of a renormalization constant: 
\begin{equation} \label{CC}
	z_L = c_L + \sum_{i}^{L} d_i(\gamma) \log(\hat{p})^i + F(\hat{p}) \;\;\;\;\;\;\;
\left(\hat{p} = p a \right).
\end{equation}
The first (finite) term is the one we are interested in (in the continuum limit). $\log$'s coefficients are known function of anomalous dimensions while the remainder is a scalar function of hypercubic invariants. The way to extract the continuum limit from \emph{hypercubic-symmetric} Taylor expansions is explained in \cite{EnzoLAT05}, which also contains a practical example, namely for the quark field renormalization constant. 

It is useful to see how our master formula Eq.~(\ref{master}) reads in a specific case, \emph{i.e.} for the  scalar current ( first order\footnote{We slightly change notation with respect to Eq.~(\ref{master}): now the order is understood as the superscript, while subscript denotes the observable.})
\begin{equation}
	\left( 1 - {{z_q^{(1)}}\over{\beta}} + \ldots \right) \left( 1 - {{z_s^{(1)}- \gamma_s^{(1)} 
\log(\hat{p}^2)}\over{\beta}} + \ldots \right) \left( 1 - {{o_s^{(1)}}\over{\beta}} + \ldots \right) = 1
\end{equation}
which determines
\begin{equation} \label{ZS1L}
	z_s^{(1)} = z_q^{(1)} - \left( o_s^{(1)} - \gamma_s^{(1)} 
\log(\hat{p}^2)  \right)
\end{equation}
Notice that $o_s^{(1)}$ is what is actually coming from measurements, while $z_q^{(1)}$ comes from an independent computation (there is no $\log$ because we are in Landau gauge). Higher orders simply requires some more algebra in order to solve for the unknown quantities (modulo some remarks we will make later).

%
\section{$Z_p/Z_s$ to four loops}
%
\begin{figure}[t]
  \begin{center} 
		\includegraphics[scale=0.6]{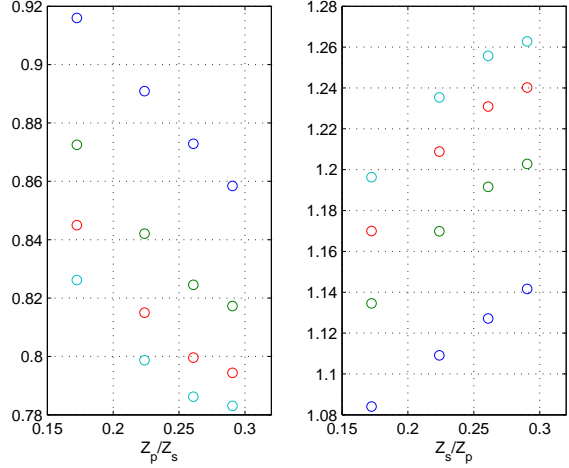}
    \caption{The resummation of $Z_p/Z_s$ and of $Z_s/Z_p$ ($n_f=2$) at $\beta=5.8$. On the $x$-axis the values of the different couplings $x_0,x_1,x_2,x_3$ (see text). Color code for $y$-axis as follows: blue is one loop result, green two loops, red three loops and pale blue four loops.}
   \label{Fig.1}
  \end{center}
\end{figure}
%
The \emph{perfect} quantity to compute is the ratio $Z_p/Z_s$. Quark field renormalization constants drops out in the ratio. There is no anomalous dimension around and so there is no conceptual limitation to our computations (apart from practical ones, at moment mainly dictated by computer memory limitations). As an extra \emph{bonus}, in our perturbative approach $Z_p/Z_s$ and $Z_s/Z_p$ come from \emph{different} signals, so that one has the handle of verifying that the series are actually inverse of each other. 

For $n_f=2$ we obtain
\begin{eqnarray}
	Z_p/Z_s & = & 1 - 0.487(2) \, \beta^{-1} - 1.46(2) \, \beta^{-2} - 5.36(7) \, \beta^{-3} + \ldots \nonumber \\
	Z_s/Z_p & = & 1 + 0.487(2) \, \beta^{-1} + 1.70(2) \, \beta^{-2} + 6.9(1) \, \beta^{-3} + \ldots 
\end{eqnarray}

One can verify that series are actually inverse of each other to a very good precision and we point out that finite size effects have been proved to be well under control by comparing results on different lattice sizes. These results come from the (quite huge) collection of three loops configurations that we stored. We actually pushed the computation even to four loops, even though with less statistics. At four loop we only quote the resummed result. For this ratio the computation is anyway safe. For quantities involving anomalous dimensions things are more cumbersome: one has also to fit an unknown $\gamma$, which comes from $\log$'s. 
Next step is now to resum the series. We now make use of BPT and resum (for $ap=1$) using different expansion parameters according to the definitions
\begin{equation}
	x_0 \equiv \beta^{-1} \;\;\;\; x_1 \equiv \frac{\beta^{-1}}{\sqrt{P}}  
\;\;\;\; x_2 \equiv - \frac{1}{2} \log(P)  \;\;\;\; x_3 \equiv \frac{\beta^{-1}}{P}
\end{equation}
where $P$ is the plaquette\footnote{While $x_2$ and $x_3$ are widely used, $x_1$ was defined just in order to better point out the general picture that emerges from the choice of different couplings: see later.}. We resum at $\beta=5.8$ in order to make contact with the results of \cite{CecVittLAT05}. Resummation is plotted in Figure 1: different colors are for different orders of resummation. We regret that there was an error in the four loop resummation shown at the conference. We here display the correct four loop result, which does not change the overall picture.

From the figure one can better understand the effect of one loop BPT: one is actually sitting on a straight line, whose slope is dictated by the first loop coefficient. Only if higher loops are taken into account a reliable understanding of what is going on can be achieved. In particular, one can easily force the series to oscillate widely by changing the coupling. We stress that convergence properties should be assessed on a case by case basis, \emph{i.e.} depending on the observable. The same holds for TI-BPT and in general if also the scale is changed while switching to a new coupling. 

As expected, at a fixed coupling we get milder and milder variations by switching on higher and higher orders, while at a fixed order we get milder and milder variations by changing the coupling. Convergence properties of the series seem reasonable for $x_2$ and $x_3$. For these couplings it is interesting to point out that if one adds to the result at a given order the deviation from the immediately lower order, one always ends up at the same result, which for example in the case of $Z_p/Z_s$ is $0.77(1)$\footnote{This is a popular way to pin down a truncation error.}. Besides that, resummed series are almost inverse of each other. The comparison with the non perturbative result in \cite{CecVittLAT05} yields encouraging result: things are working pretty well in the \emph{RI'-MOM} scheme. 

%
\begin{figure}[t]
  \begin{center} 
		\includegraphics[scale=0.5]{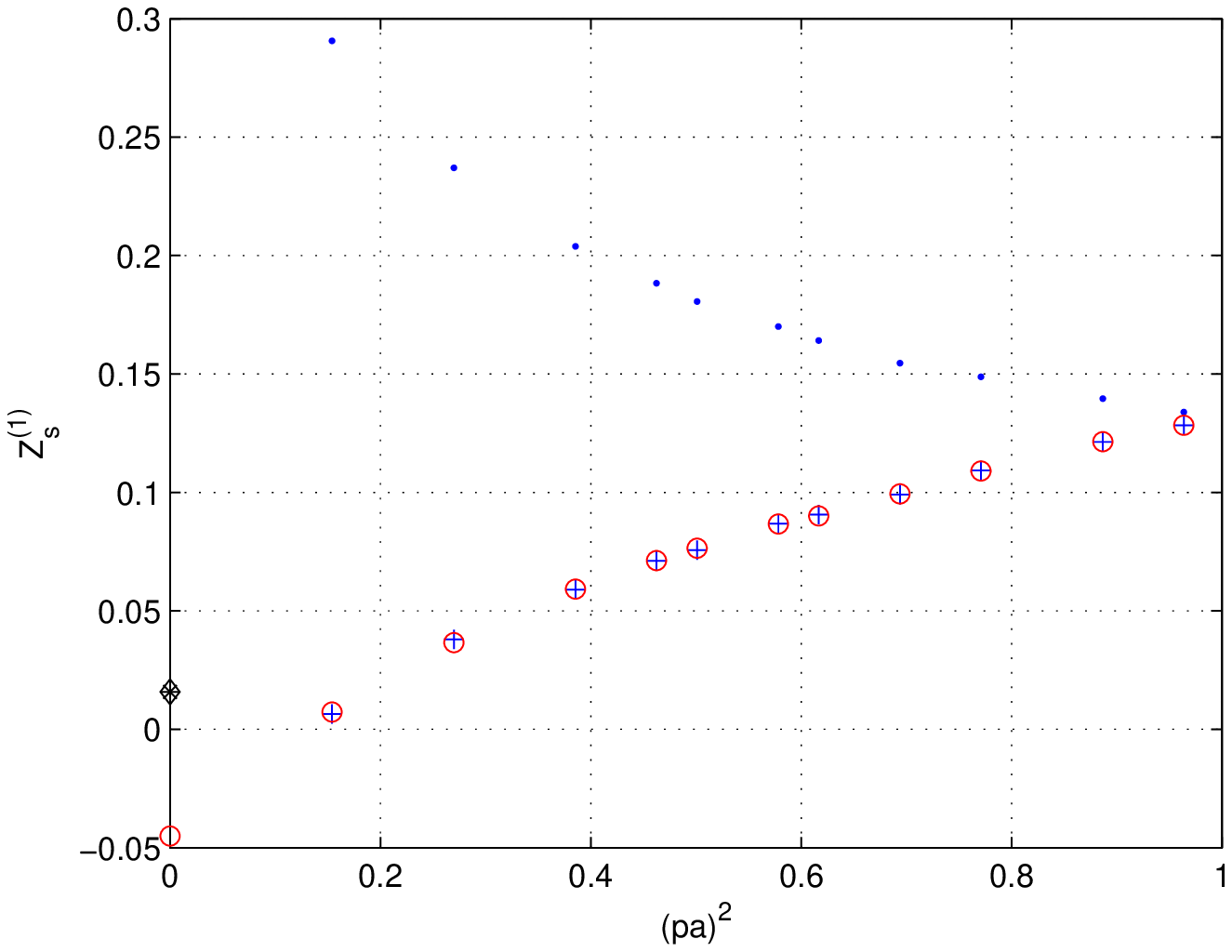}
		\includegraphics[scale=0.5]{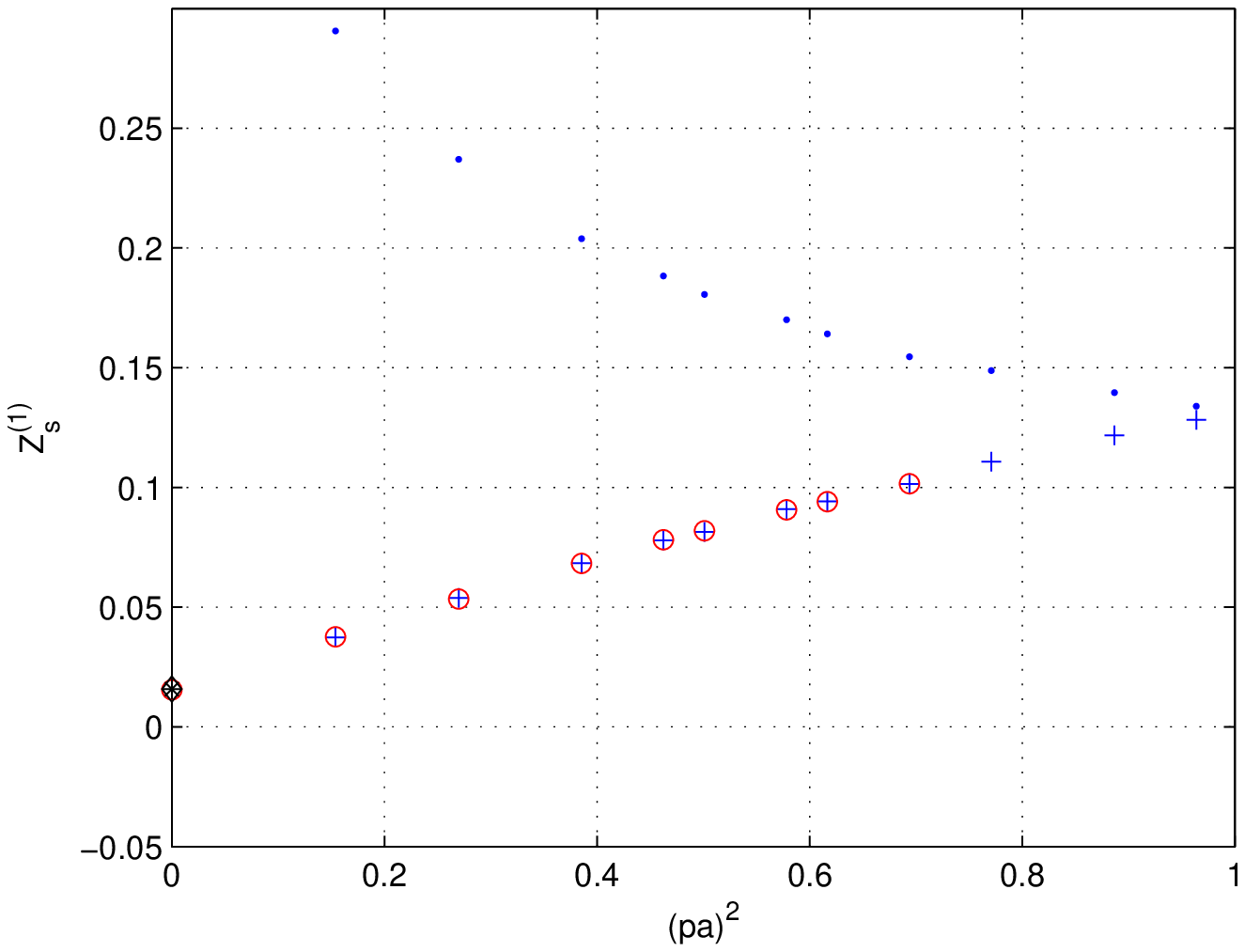}
    \caption{Computation of one loop renormalization constant for the scalar current. Upper points are the unsubtracted $o_s^{(1)}$, while lower (ticker symbols) stand for the subtracted $o_s^{(1)} - \gamma_s^{(1)} \log(\hat{p}^2)$. Analytic result is marked with a darker symbol. On the left: no correction for finite volume. On the right: finite volume \emph{tamed-log} taken into account.}
   \label{Fig.1}
  \end{center}
\end{figure}%

%
\section{Treating anomalous dimensions}

Let us go back to Eq.~(\ref{ZS1L}) (the scalar current at one loop). We make use of this one loop example since the analytic result is known \cite{MartiZ}. A naive application of Eq.~(\ref{ZS1L}) yields what is plotted in Figure 2 on the left: the result is missed, \emph{as if one were subtracting too much by subtracting the log}. The effect is systematic: a similar picture is obtained for example for $Z_p$. By computing on different lattice sizes, one realizes that is a finite size effect, getting worse and worse as the volume decreases. We stress that this is a $pL = = {{\hat{p}}\over{a}} N a = \hat{p} N$ effect, $L$ being expressed in terms of the lattice spacing as $L=Na$. One could say that the $\log$ is \emph{tamed} by finite volume. Since we want to regard our finite volume  computations as inifite volume approximations, we need to correct for this effect. A solution comes from computing the expected $\log$ in the continuum at the same values of $pL=\hat{p} N$ one is interested in. The result is shown in Figure 2 on the right. Notice anyway that if one stays away from the lower momenta results are anyway quite safe.

%
\section{Conclusions and prospects}

NSPT is a valuable tool to compute logarithmic-divergent renormalization constants for lattice QCD. Now that configurations are stored, a lot of computations are possible and, in particular, a full account of quark bilinears to three loops will be ready soon. The main caveat when making use of our method is that care is needed when dealing with  $\log$'s, because of finite volume effects. 

We are in a position to assess convergence properties of the series and to shed some light on BPT, a blind application of which is never advisable. Even when a non perturbative result is available, our method can provide an important check to better assess systematic errors.

%

\end{document}